\documentclass[11pt]{article}

\usepackage[margin=1in]{geometry}
\usepackage[T1]{fontenc}
\usepackage[utf8]{inputenc}
\usepackage{lmodern}
\usepackage{microtype}
\usepackage{hyperref}
\usepackage{enumitem}
\usepackage{booktabs}
\usepackage{array}
\usepackage{graphicx}
\usepackage{xcolor}

\hypersetup{
  colorlinks=true,
  linkcolor=blue!50!black,
  urlcolor=blue!50!black,
  citecolor=blue!50!black,
  pdftitle={ResearchPilot: A Local-First Multi-Agent System for Literature Synthesis and Related Work Drafting},
  pdfauthor={Peng Zhang}
}

\title{\textbf{ResearchPilot: A Local-First Multi-Agent System for Literature Synthesis and Related Work Drafting}}
\author{Peng Zhang\\\texttt{pengz7720@gmail.com}}
\date{}

\begin{document}
\maketitle

\begin{abstract}
ResearchPilot is an open-source, self-hostable multi-agent system for literature-review assistance. Given a natural-language research question, it retrieves papers from Semantic Scholar and arXiv, extracts structured findings from paper abstracts, synthesizes cross-paper patterns, and drafts a citation-aware related-work section. The system combines FastAPI, Next.js, DSPy, SQLite, and Qdrant in a local-first architecture that supports bring-your-own-key model access and remote-or-local embeddings. This paper describes the system design, typed agent interfaces, persistence and history-search mechanisms, and the engineering tradeoffs involved in building a transparent research assistant. Rather than claiming algorithmic novelty, we present ResearchPilot as a systems contribution and evaluate it through automated tests and end-to-end local runs. We discuss limitations including external API rate limits, abstract-only extraction, incomplete corpus coverage, and the lack of citation verification.
\end{abstract}

\section{Introduction}

Literature review is one of the most time-intensive stages of research. A researcher entering a new area must identify relevant work, understand the methods and datasets used in prior papers, compare findings across studies, and organize those findings into a coherent narrative. These subtasks are conceptually distinct, but they are often performed in a single manual loop of search, reading, note-taking, and writing.

Large language models (LLMs) are useful for summarization, information extraction, and drafting, but one-shot prompting is an awkward fit for literature review. A single prompt asked to ``summarize the literature'' tends to blur retrieval and reasoning, obscure intermediate evidence, and make it difficult to inspect the provenance of a synthesis claim. It also encourages systems to behave like opaque answer generators rather than transparent assistants.

ResearchPilot takes a different approach. It decomposes literature-review assistance into a fixed sequence of four stages: paper retrieval, per-paper extraction, cross-paper synthesis, and related-work drafting. Each stage is implemented as a constrained module with explicit inputs and outputs, and the full pipeline is exposed through a local-first web application with streaming status updates and report persistence.

This paper makes four contributions. First, it presents a four-stage multi-agent architecture for literature retrieval, structured abstract extraction, cross-paper synthesis, and related-work drafting. Second, it shows how DSPy can be used as an orchestration layer with typed input/output contracts while still supporting multiple model providers through configuration rather than prompt rewrites. Third, it describes a self-hostable full-stack implementation built with FastAPI, Next.js, SQLite, and Qdrant, including report history and semantic search over prior runs. Fourth, it documents the engineering constraints that arise in such a system, including API rate limits, structured-output failures, embedding portability, and abstract-only fidelity limits.

We do not claim a novel learning algorithm or benchmark-leading performance. The contribution is a transparent, inspectable system design and a reproducible workflow for AI-assisted literature synthesis.

\section{Problem Setting}

The system takes as input a natural-language research question such as ``What methods improve factuality in long-form LLM generation?'' The desired output is not a single free-form answer but a bundle of intermediate and final artifacts: a retrieved set of papers with abstracts and metadata, a structured extraction for each paper, a synthesis of consensus, contradictions, and open gaps across papers, and a citation-aware related-work draft in markdown.

This task framing intentionally constrains the system. ResearchPilot does not attempt to ingest full papers by default, verify every citation claim, or provide exhaustive domain coverage. Its goal is narrower: provide a transparent first-pass research workflow that helps a user move from a question to a structured draft while preserving inspectable intermediate outputs.

\section{Related Work}

\subsection{Retrieval-Augmented Generation}

Retrieval-Augmented Generation (RAG) augments language-model generation with retrieved external context to improve grounding and expand coverage beyond training data~\cite{lewis2020rag}. ResearchPilot is adjacent to this paradigm, but it inserts structured extraction and synthesis stages between retrieval and generation instead of feeding retrieved documents directly into a final prompt.

\subsection{Multi-Agent LLM Systems}

Recent LLM systems have explored decomposing tasks across multiple agents or roles. AutoGPT~\cite{autogpt2023} and MetaGPT~\cite{hong2023metagpt} are representative examples of multi-component workflows that divide planning and execution responsibilities. ResearchPilot also decomposes a complex task, but it uses a fixed sequential pipeline rather than planner-style or dynamically selected agents. This choice reflects the relatively stable structure of literature-review work.

\subsection{LLM Orchestration Frameworks}

LangChain~\cite{langchain2022} and LlamaIndex~\cite{llamaindex2022} provide general abstractions for chaining LLM calls, tools, and retrieval components. DSPy~\cite{khattab2023dspy} emphasizes declarative signatures and typed field descriptions. ResearchPilot uses DSPy because the agent boundaries naturally map to structured input and output contracts.

\subsection{Scholarly Search and Review Tools}

Semantic Scholar~\cite{kinney2023semanticscholar} and arXiv provide accessible scholarly search substrates. Hosted literature-review tools further illustrate growing demand for AI-supported literature analysis, but many such systems do not expose intermediate reasoning stages or programmable structured outputs. ResearchPilot is designed instead as an open, inspectable, self-hostable system.

\section{System Overview}

ResearchPilot consists of four layers: a Next.js frontend for question entry, runtime configuration, live status viewing, report display, and history search; a FastAPI backend exposing pipeline and report APIs; a DSPy-orchestrated four-agent pipeline; and a persistence layer combining SQLite and Qdrant.

\begin{figure}[t]
\centering
\includegraphics[width=\linewidth]{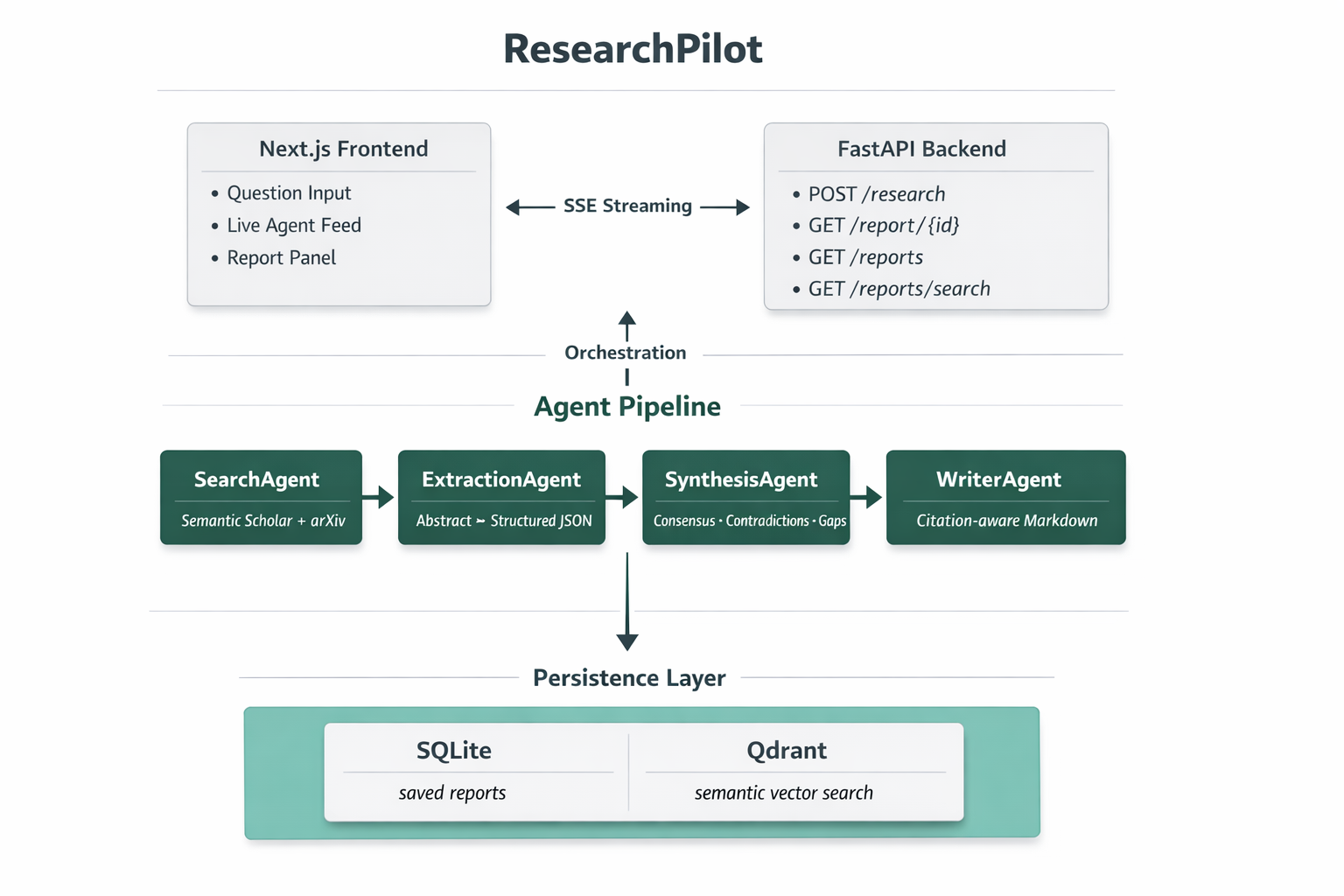}
\caption{Architecture of ResearchPilot. The frontend communicates with the FastAPI backend, which orchestrates a four-stage agent pipeline and persists reports in SQLite and Qdrant.}
\label{fig:architecture}
\end{figure}

The frontend submits a research question and optional runtime overrides to the backend. The backend launches the pipeline, streams lifecycle events during execution, persists the completed report, and exposes prior runs through report-history endpoints. Qdrant is used both to store vectorized artifacts and to support semantic search over prior reports.

\section{Agent Pipeline}

\subsection{SearchAgent}

\texttt{SearchAgent} queries Semantic Scholar and arXiv asynchronously. The implementation filters out entries without abstracts, deduplicates by DOI or title, and truncates the final merged list to at most ten papers. Search is designed to tolerate partial source failure: if one source fails, for example due to HTTP 429 rate limiting, the other source can still provide a usable paper set. Warnings are surfaced to both the event stream and the final report.

The output is a list of paper objects with fields including identifier, title, abstract, source, URL, year, authors, and optional DOI.

\subsection{ExtractionAgent}

\texttt{ExtractionAgent} processes each abstract independently using a DSPy signature. The model is asked to return JSON with five arrays: \texttt{claims}, \texttt{methods}, \texttt{datasets}, \texttt{results}, and \texttt{limitations}.

This stage runs once per paper and produces a \texttt{PaperExtraction} object. Running extraction per paper keeps context windows small and enables granular progress reporting. The current implementation is sequential rather than parallelized.

\subsection{SynthesisAgent}

\texttt{SynthesisAgent} consumes the list of extraction objects and produces a cross-paper synthesis with three fields: \texttt{consensus}, \texttt{contradictions}, and \texttt{open\_gaps}.

This stage is responsible for consolidating recurring patterns and disagreements across the extracted paper summaries rather than summarizing each paper independently.

\subsection{WriterAgent}

\texttt{WriterAgent} takes the synthesis object and a bibliography derived from the retrieved papers and generates a markdown related-work draft with inline labels such as \texttt{[R1]} and \texttt{[R2]}. The generated output includes a references section that maps labels to the retrieved paper list.

The writer is intended as a first-draft assistant. It does not verify that every citation label is perfectly supported by the referenced paper.

\subsection{Stage Contracts}

\begin{table}[t]
\centering
\small
\begin{tabular}{>{\raggedright\arraybackslash}p{0.17\linewidth} >{\raggedright\arraybackslash}p{0.31\linewidth} >{\raggedright\arraybackslash}p{0.39\linewidth}}
\toprule
Stage & Input & Output \\
\midrule
SearchAgent & research question & list of papers with abstracts and metadata \\
ExtractionAgent & question + single paper abstract & structured extraction with claims, methods, datasets, results, limitations \\
SynthesisAgent & question + list of extractions & consensus, contradictions, open gaps \\
WriterAgent & question + synthesis + bibliography & markdown related-work draft with inline reference labels \\
\bottomrule
\end{tabular}
\caption{Pipeline stages and their primary input/output contracts.}
\label{tab:stages}
\end{table}

\section{Orchestration and Interface Design}

\subsection{DSPy Signatures}

Each agent is implemented as a DSPy module or signature with explicit input and output fields. This design keeps the prompts structured around data contracts rather than large hand-written prompt templates. It also supports provider changes by swapping environment configuration or per-run runtime overrides.

\subsection{Streaming Research API}

The backend exposes a \texttt{POST /research} endpoint that returns a streaming response. During execution, the stream emits lifecycle events such as \texttt{queued}, \texttt{agent\_started}, \texttt{agent\_progress}, \texttt{agent\_completed}, \texttt{done}, and \texttt{error}.

The current frontend consumes this stream using an incremental \texttt{fetch()} reader over a POST request rather than the browser's native \texttt{EventSource} interface. This choice allows the request body to include both the research question and runtime configuration overrides.

\subsection{Frontend Experience}

The single-page frontend renders the live event feed, final markdown output, synthesized sections, warnings, retrieved papers, and report-history results. It also supports per-run overrides of provider, model, API key, base URL, and embedding mode without requiring a backend restart.

\section{Persistence and Semantic History Search}

Completed reports are stored in SQLite. Each saved report includes the original question, retrieved papers, structured extractions, synthesis output, markdown draft, warnings, references, and creation timestamp.

Qdrant is used as a vector-backed artifact store. Report vectors support semantic search over prior runs, while paper and extraction artifacts are also persisted for future retrieval-oriented extensions. When vector search is unavailable, the backend falls back to simple keyword matching over report questions.

Embeddings can be produced in three modes: \texttt{remote}, which uses a provider-hosted embedding API; \texttt{local}, which uses a local open-source embedding model; and \texttt{auto}, which prefers remote embeddings when configured and falls back to local embeddings otherwise.

The default local model is \texttt{BAAI/bge-small-en-v1.5}, which makes the system usable even when the chosen LLM provider does not supply a convenient embedding API.

\section{Implementation}

The backend is implemented in Python 3.11+ using FastAPI, DSPy, \texttt{httpx}, SQLite, and the Qdrant Python client. The frontend is implemented in TypeScript with Next.js App Router and React. The overall system is packaged with Docker Compose and includes a Makefile for local startup, logs, and tests.

Model access is designed around bring-your-own credentials. The backend reads a local environment file, and the frontend can override runtime model settings per request. Supported usage patterns include OpenAI-compatible APIs, Anthropic, Groq, OpenRouter, and related endpoints configured through provider/model/base-URL fields.

ResearchPilot is intentionally local-first. It is designed to be run by users on their own machines or infrastructure instead of depending on a centrally hosted service.

\section{Evaluation}

We treat ResearchPilot as a systems prototype rather than a benchmarked research model. Accordingly, the evaluation emphasizes implementation correctness, end-to-end functionality, and observed failure modes.

\subsection{Automated Validation}

The codebase includes backend tests and a frontend smoke test. These automated checks cover successful streaming and report lookup through the backend API, propagation of setup failures as streamed error events, report listing and keyword fallback for report search, safe exposure of runtime configuration without leaking API keys, deduplication and partial-failure behavior in paper search, embedding-mode selection and local embedding behavior, and basic frontend rendering of the workflow shell.

In the current codebase, this corresponds to 11 backend tests and 1 frontend test passing locally. These tests do not establish output quality, but they do validate core system behaviors and error-handling paths.

\subsection{Manual End-to-End Validation}

The system was also exercised through local end-to-end runs in the web interface. These checks focused on whether a research question can move through the full pipeline, whether agent lifecycle events are visible to the user, whether the final report is persisted and reloadable, and whether history search returns prior reports.

Representative example prompts used during development included questions about retrieval-augmented generation, long-form factuality, diffusion models for time-series forecasting, and graph neural networks for molecular-property prediction. These runs were used as functional validation and debugging inputs rather than as a formal benchmark.

On March 14, 2026, we recorded a small local evaluation using the deployed application with the Groq-hosted \texttt{llama-3.3-70b-versatile} model and local embeddings. One end-to-end run on the query ``What are the recent trends in retrieval-augmented generation for question answering?'' completed successfully in 12.47 seconds, returning 10 papers, 10 structured extractions, a synthesis with 4 consensus items, 2 contradictions, and 3 open gaps, and a related-work draft of 2046 characters. A warning was emitted because Semantic Scholar returned HTTP 429, but the pipeline still completed using arXiv results.

\begin{table}[t]
\centering
\small
\begin{tabular}{>{\raggedright\arraybackslash}p{0.34\linewidth} c c c >{\raggedright\arraybackslash}p{0.2\linewidth}}
\toprule
Query & Completed & Time (s) & Papers & Main issue \\
\midrule
Recent trends in RAG for question answering & Yes & 12.47 & 10 & Semantic Scholar 429 warning \\
Methods improving factuality in long-form LLM generation & No & 28.42 & -- & provider token rate limit \\
Main limitations of GNNs for molecular property prediction & No & 29.24 & -- & provider token rate limit \\
\bottomrule
\end{tabular}
\caption{Small local end-to-end evaluation performed on March 14, 2026. Results are reported as functional system observations, not as a comparative benchmark.}
\label{tab:eval}
\end{table}

We also executed a subsequent three-query batch on the same day to test repeat-run behavior. In that batch, repeated requests terminated with provider-side token rate-limit errors before full completion. This is an important operational result: even when the pipeline implementation is stable, end-to-end reliability remains constrained by external model-provider quotas.

\subsection{Observed Failure Modes}

The manual evaluation surfaced several recurring failure modes. External API rate limits remain a practical issue, as Semantic Scholar may return HTTP 429 without an API key or under bursty access. Provider-specific JSON variance also matters: some models occasionally omit fields in extraction output, making schema validation and defensive parsing necessary. A third issue is embedding-provider mismatch, since not all model providers expose embeddings in the same way, which motivated a local fallback path. Finally, infrastructure components such as streaming and vector storage can fail independently, so the pipeline benefits from non-fatal warning paths and fallback logic.

\begin{table}[t]
\centering
\small
\begin{tabular}{>{\raggedright\arraybackslash}p{0.27\linewidth} >{\raggedright\arraybackslash}p{0.28\linewidth} >{\raggedright\arraybackslash}p{0.33\linewidth}}
\toprule
Failure mode & Example effect & Current mitigation \\
\midrule
Semantic Scholar rate limiting & search warnings or single-source retrieval & partial-failure tolerance and optional API key support \\
Malformed model JSON & extraction or synthesis parsing failure & schema validation and defensive parsing \\
Missing remote embeddings & vector search unavailable for some providers & local embedding fallback and keyword search fallback \\
Persistence-layer errors & Qdrant write or lookup failures & warning surfacing and SQLite-backed report persistence \\
\bottomrule
\end{tabular}
\caption{Observed failure modes and current mitigations in the implementation.}
\label{tab:failures}
\end{table}

\section{Limitations and Future Work}

ResearchPilot has several important limitations.

First, retrieval is currently limited to Semantic Scholar and arXiv. This is enough for a broad prototype but not for many domain-specific workflows, especially in medicine, engineering, or venues concentrated in closed digital libraries.

Second, the system works from abstracts rather than full papers. Abstracts are often insufficient for precise reporting of datasets, experimental settings, or quantitative outcomes.

Third, the generated related-work draft is citation-aware but not citation-verified. The system tracks reference labels and retrieved sources, but it does not independently prove that every generated sentence is fully supported by the cited paper.

Fourth, the current evaluation is functional rather than comparative. A stronger future study would compare the multi-stage pipeline against simpler baselines, include human ratings for output usefulness and citation quality, and measure stage-level latency and completion rates across a larger evaluation set.

Promising future directions include full-text PDF ingestion, domain-specific retrieval connectors, bounded-concurrency extraction, claim or citation verification agents, and reuse of prior report artifacts as retrieval context for new runs.

\section{Conclusion}

ResearchPilot demonstrates that literature-review assistance can be implemented as a transparent multi-stage pipeline instead of a single opaque prompt. By decomposing search, extraction, synthesis, and drafting into explicit modules with typed interfaces, the system becomes easier to inspect, debug, persist, and extend.

The main contribution is not a new learning method but a practical open-source architecture for local-first research assistance. For researchers and developers who value inspectability and self-hostability, this architecture offers a useful baseline for future work on AI-assisted literature analysis.

\end{document}